\documentclass[%
preprint,
amsmath,amssymb,
aps,onecolumn
]{revtex4-1}

\usepackage{graphicx}
\usepackage{dcolumn}
\usepackage{bm}
\usepackage{braket}
\usepackage{hyperref}
\usepackage{subcaption}
\usepackage{amsmath}
\usepackage{amssymb}   
\usepackage{wrapfig}
\usepackage{array}
\usepackage{caption}
\usepackage{blindtext}
\usepackage{color}
\usepackage{times, verbatim}
\usepackage{pstricks}
\usepackage{caption}
\begin{document}


\title{Monte Carlo studies of modified scalable designs for quantum computation}

\author{Guan-Hao Feng}
\author{Jingwei Chen}
\author{Hong-Hao Zhang}
\email{zhh98@mail.sysu.edu.cn}

\affiliation{School of Physics, Sun Yat-sen
University, Guangzhou 510275, China}


\begin{abstract}
As the building blocks of topological quantum computation, Majorana zero modes (MZMs) have attracted tremendous attention in recent years. Scalable mesoscopic island designs with MZMs show great potential in quantum information processing. However, these systems are susceptible to quasi-particle poisoning which would induce various parity-breaking errors. To solve this problem, we modify the mesoscopic islands with gate-tunable valves and non-topological backbones. We study the lifetime of the Majorana qubits on these modified islands which are coupled to local bosonic and fermionic thermal baths. We consider both the parity-breaking and parity-preserving errors, and propose a parity correction scheme. By using Jordan-Wigner transformation, we analyze the probability of logical X and Y errors. The open quantum system is described by the Pauli master equation, and standard Monte Carlo simulations are applied to observe the behavior of the system when the parity correction proposal is implemented. The results demonstrate that (1) our parity correction proposal is effective to most of the parity-breaking errors; (2) the lifetime of the qubit benefits from larger island size before it meets the threshold; (3) small chemical potential $ \mu $ on the non-topological backbones and fine tuned paring potential $ \Delta $ of the topological bulk segment are required for high probability of correctness. Our results provide an effective error correction scheme for the parity-breaking errors.
\end{abstract}

\maketitle
\tableofcontents

\section{\label{sec:level1}Introduction}
MZMs in one-dimensional semiconductor-superconductor are candidates to achieve topological quantum computing (TQC)  \cite{kitaev2003fault,RevModPhys.80.1083,PhysRevX.4.011036,PhysRevLett.104.040502,RN59,lutchyn2018majorana}.
The non-abelian statistics and spatial separation of MZMs are beneficial for Fault-torrent topological quantum computing (FTQC) \cite{PhysRevB.81.125318, alicea2011non,PhysRevB.100.014511,PhysRevA.77.012327}.
As quantum gates are implemented by the manipulations of the non-local MZMs, and the quantum information stored in the topological states are expected to resist against any local perturbation intrinsically \cite{PhysRevA.51.1015,PhysRevB.99.144521}. However, TQC could be still susceptible to the perturbations involving MZM operators \cite{PhysRevB.85.121405,Bravyi2012}.

FTQC is expected to provide long-lifetime Majorana-based qubits, and thus it provides great potential in physics and material science \cite{RN64,He294,Mourik1003,PhysRevB.98.155414,PhysRevA.90.012321,PhysRevX.7.021011,PhysRevLett.117.076403}. Encoding Majorana qubits is a key step in FTQC \cite{PhysRevX.4.011051,PhysRevX.7.031048,PhysRevB.88.075431,PhysRevA.57.127}. For a robust logical qubit, it usually consists of several physical qubits. That is, FTQC is actually based on encoding of large-scale qubits.

For the hardware of FTQC, the semiconductor nanowire combining s-wave superconductor can realize effective spinless p-wave superconductor \cite{kitaev2001unpaired,LIEB1961407,doi:10.1021/nl303758w,PhysRevB.87.241401}. With the use of this heterostructure, proposals for topological quantum memories and Majorana-based quantum computing are appealing for both experimental and theoretical research \cite{PhysRevB.96.060505, PhysRevB.94.094303}. It is necessary to combine the large-scale qubits encoding and the hardware design \cite{PhysRevX.7.031048}.

Recently, scalable network designs for FTQC set up a platform to realize a large scale of Majorana qubits manipulation \cite{PhysRevB.95.235305}. Following Ref.~\cite{aasen2016milestones}, gate-tunable valves are used to manipulate and readout the MZMs. Although the Majorana qubits are topologically protected in these network designs, the quasi-particle poisoning (QPP) induced by environment perturbation can arise various topological fermion parity-breaking errors in the system. Based on these designs, the encoding of Majorana qubits can substantially decrease the probability of these errors. The error correcting codes have been studied in detail, such as topological surface codes and color codes wildly used in quantum computation \cite{PhysRevB.97.205404,PhysRevX.5.041038,PhysRevA.88.062313,PhysRevA.54.3824,PhysRevX.2.041021}. These codes work well in correcting topological fermion parity-preserving errors. However, for the parity breaking errors, it is necessary to correct the topological fermion parity before the application of the implementation of the decoding.

 In this paper, we modify the network designs in Ref.~\cite{PhysRevB.95.235305} with gate-tunable valves and semiconductor-superconductor backbone which is regarded as non-topological spinless p-wave superconductor. In order to study the lifetime of the Majorana qubits, we encode the system with Kitaev's 1D toric code and simulate the system coupling to local thermal environment. Both parity-breaking and parity-preserving circumstances are taken into consideration. The error excitations induced by the thermal environment can be detected and corrected by the measurements of error syndrome \cite{PhysRevB.92.115441,PhysRevA.86.052340}. We also suggest a proposal to correct the parity-breaking errors. The gate-tunable valves are used to switch the error excitation state to the charge state, and switch the topological fermion parity to charge parity which is detectable as the experiment proposal in Ref.~\cite{RN62}. Therefore, it is possible to correct the parity errors caused by QPP. We firstly consider the island configuration which contains 2 MZMs and couples to one quantum dot. The length of the system is modified to observe the impact on the probability of error occurrence. Then similar discussion is implemented in 4-MZM island.
The validity of our error correction scheme is verified by Monte Carlo simulation.

This paper is organized as follows. In Sec.~\ref{sec:Kitaev}, we briefly review the concept of topological and nontopological states in Kitaev's 1D toy model. In Sec.~\ref{sec:topologicalcooperpair}, we describe the modified hexon, and compare the differences from Ref.~\cite{PhysRevB.95.235305}. Furthermore, the box representation is implemented in this model \cite{PhysRevB.92.115441}. In Sec.~\ref{sec:majorana}, we encode the system with Kitaev's 1D toric code and introduce the bosonic and fermionic local thermal baths. We describe the open system with Pauli master equation. In Sec.~\ref{sec:errorcorrection}, we suggest our parity correcting proposal and make use of the error correction algorithm in Ref.~\cite{wootton2015simple} to discuss the probability of correctness. With the use of Jordan-Wigner transformation, we analyze the probability of logical X and Y errors. In Sec.~\ref{sec:montecarlo}, we implement the Monte Carlo simulation on the 2-MZM and 4-MZM islands respectively. Finally, in Sec.~\ref{sec:conclusion}, we summarize our analytical and Monte Carlo simulation results.

\section{Modified scalable island with MZMs}
The network layouts of one dimensional topological superconductors are the basis of FTQC. In this section, we briefly recall one dimensional Kitaev's toy model and topological superconductor. According to the scalable network layouts in Ref. \cite{PhysRevB.95.235305}, we propose the modified topological cooper pair box and discuss the its property coupling to the thermal baths.
\subsection{\label{sec:Kitaev}Kitaev chain and topological superconductor}
 One dimensional hybrid nanowire combining semiconductor and superconductor materials is a proposal to obtain MZMs. Without loss of generality, one can show that the Hamiltonian of such a hybrid nanowire is equivalent to that of a spinless p-wave superconductor in low energy limit \cite{hansen2014unpaired}, which is described by the 1D Kitaev toy model in real space \cite{kitaev2001unpaired}
\begin{eqnarray}
H
=-\mu\sum_{i=1}^N\left(c_i^\dagger c_i-\frac{1}{2}\right)+\sum_{i=1}^{N-1}\left(-tc_{i}^\dagger c_{i+1}+\Delta c_ic_{i+1}+\mathrm{h.c.}\right),
\end{eqnarray}
where $  \Delta=|\Delta|e^{i\phi}  $ is the pairing potential of the bulk segments, $ \mu $ is the chemical potential, $ t $ is the tunneling strength. It is useful to rewrite the Hamiltonian in terms of Majorana operators $ \gamma_{2i-1}=e^{i\phi/2}c_i+e^{-i\phi/2}c_i^\dagger $ and $ \gamma_{2i}=-i(e^{i\phi/2}c_i-e^{-i\phi/2}c_i^\dagger) $.
\begin{equation}\label{key}
H=-\frac{i}{2}\mu\sum_{i}^N\gamma_{2i-1}\gamma_{2i}
+\frac{i}{2}\sum_{i}^{N-1}\bigg[(|\Delta|+t)\gamma_{2i}\gamma_{2i+1}+(|\Delta|-t)\gamma_{2i-1}\gamma_{2i+2}\bigg]
\end{equation}
There are two phases in such a hybrid nanowire, i.e. the topological phase and non-topological phase \cite{PhysRevB.92.115441}.These Hamiltonians can respectively be written as
\begin{eqnarray}
H_\text{top}&=&i|\Delta|\sum_{i=1}^{N-1}\gamma_{2i}\gamma_{2i+1}, \quad \text{for}~\mu_{j}=0, t=|\Delta|\\
H_\text{nontop}&=&-\frac{i}{2}\sum_{i=1}^{N}\mu\gamma_{2i-1}\gamma_{2i}, \quad \text{for}~\mu_{j}< 0, t=|\Delta|=0.
\end{eqnarray}
For the topological phase, one can diagonalize the Hamiltonian by introducing the quasi-particle operators $ d_i=\frac{1}{2}(\gamma_{2i}+i\gamma_{2i+1}), i\geq 1$ and $ d_0=\frac{1}{2}(\gamma_{1}+i\gamma_{2N}) $\cite{alicea2011non}, which leads to
\begin{equation}\label{key}
H_\text{top}=\sum_{i=1}^{N-1}|\Delta|(2d_i^\dagger d_i-1).
\end{equation}
Noting that the quasi-particle $ d_0 $ is absent from $ H_\text{top} $, suggesting that there are two degenerate ground states, i.e., $ \ket{0} $ and $ \ket{1}=d_0^\dagger\ket{0} $,thus $ \gamma_{1} $ and $ \gamma_{2N} $ are isolated MZMs and arise the two degenerate ground states $ \ket{0} $ and $ \ket{1}=d_0^\dagger\ket{0} $. We can represent the two degenerate ground states of the whole hybrid nanowire as
\begin{eqnarray}\label{key}
&&\ket{0}_\text{wire}=\ket{0...0},\\
&&\ket{1}_\text{wire}=d_0^\dagger\ket{0...0}=\ket{1,0...0}.
\end{eqnarray}
Noting that any quasi-particle creation operators $ d_{i\neq 0}^\dagger $ can cause the excitations of the hybrid nanowire with energy $ |\Delta| $, which is regarded as QPP. The interaction between the environment and the hybrid nanowire is the primary source of QPP. In Ref. \cite{PhysRevB.95.235305}, hexon and tetron architectures have been discussed in respect of projective measurements and Clifford completeness. In this paper, we will modify the hexon and tetron architectures, and study the lifetime of Majorana code based on the scalable islands.
\subsection{\label{sec:topologicalcooperpair}Modified topological cooper pair box}
\begin{figure}[htbp]
	\centering
	\includegraphics[width=0.75\textwidth]{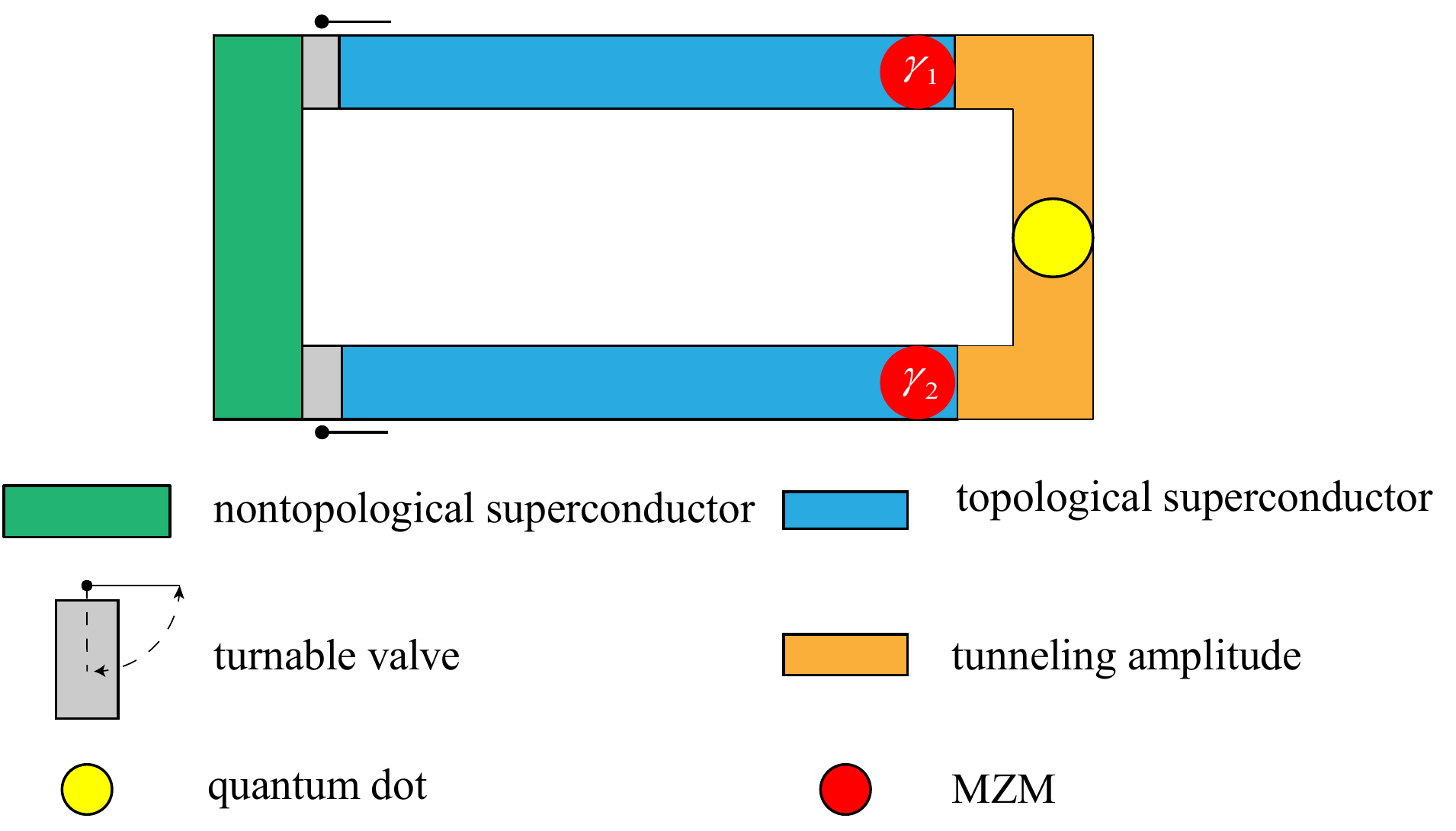}
	\caption{The minimal structure for the Majorana measurements in hexon \cite{PhysRevB.95.235305}. The parity of the structure is determined by $ i\gamma_{1}\gamma_{2} $. The corresponding logical qubits are $ \ket{0}=\ket{0_{\gamma_{1}\gamma_{2}}} $ and $ \ket{1}=\ket{1_{\gamma_{1}\gamma_{2}}} $. The two valves (gray) on the left side of the topological bulk segments is gate-tunable. When the valves are open, the Josephson energy dominates the island, i.e. $ E_J \gg E_C $, and the external magnetic field $ \vec{B} $ induces the Zeeman coupling which draws the bulk segment into topological state. When the valves are closed, i.e.$ E_J \ll E_C $, the bulk segments turn into non-degenerate charge state $ \ket{Q,P_{\text{odd}}} $ or $ \ket{Q,P_{\text{even}}} $\cite{aasen2016milestones}. }
	\label{fig:structure-1}
\end{figure}

We begin with the modified island (composed of one back bone, two bulk segments and two Josephson junctions) for measuring the two-MZM parity $ p_{12} $, which is defined as the eigenvalue of $ i\gamma_1\gamma_2 $. The system consists of an island coupling to a quantum dot. As depicted in Fig.~\ref{fig:structure-1}, two gate-tunable valves are set to connect the nontopological backbone and the two topological segments, and the main role of the valves is to turn the error excitation states into nondegenerate charge states, so that the topological parity detection and correction can be applied, we will further discuss in Section~\ref{sec:majorana}. The charging energy of the two topological segments is $ E_{C} $, and the gate-tunable valves lead to the tunable Josephson Energy $ E_J $. Consequently, the radio of $ E_J $ and $ E_{C} $ is tunable, similar configuration can be found in Ref.~\cite{aasen2016milestones}. By tuning the Josephson Energy $ E_J $ much larger than the charging energy $ E_{C} $, the topological segments would host a pair of MZMs. The island with an overall charging energy $ E_C $ can be regarded as a topological Cooper pair box. The Hamiltonian of the modified island decoupled with the quantum dot is
\begin{eqnarray}\label{key}
H_{\text{island}}&=&\sum_{\alpha=1,2}(H_{J},\alpha)+H_C+H_{BCS},
\end{eqnarray}
where $H_{\text{BCS}}=\sum_{\alpha=1,2}H_{\text{top},\alpha}+H_{\text{nontop}}$, and $ H_\text{C} $ is the Hamiltonian of overall charging energy of the island and takes the form
\begin{equation}\label{key}
H_C=E_C (\hat{N_S}-N_g)^2.
\end{equation}
Here, the operator $ \hat{N_S} $ denotes the number of Cooper pairs exceeding the neutrality in the island. Actually, the length of topological segments is much larger than the width of island, so $ \hat{N_S} $ mainly counts the charge number in topological bulk segments. Obviously, the eigenstates of $ H_C $ take the form $ \ket{N_S} $ where $ N_S=...,-2,-1,0,1,2... $. In general, the form of $H_{J,\alpha}$ can be written as
\begin{equation}\label{key}
H_{J,\alpha}=-E_{J,\alpha} \cos(\hat{\varphi_\alpha}),
\end{equation}
where $ \alpha=1,2 $ denotes the Josephson energy in the upper Josephson junction and lower Josephson junction respectively. Assuming $ E_{J,1}=E_{J,2} $ can obtain \cite{PhysRevLett.76.4408}
\begin{equation}\label{aa}
\sum_{\alpha=1,2}H_{J,\alpha}=-E_{J,1}\cos(\hat{\varphi_1})+E_{J,1}\cos(\hat{\varphi_2})\approx -E_J(\phi) \cos[\hat{\varphi}+\gamma({\phi})],
\end{equation}
where $ E_{J}(\phi)=(E_{J,1}^2+
E_{J,2}^2+2E_{J,1}E_{J,2}\cos \phi)^{1/2} $ is the overall Josephson energy, and $ \phi=\varphi_1+\varphi_2 $ is the overall Josephson phase. Noting that
$ \tan \gamma(\phi)=-[(E_{J,1}-E_{J,2})/(E_{J,1}+E_{J,2})]\tan(\phi/2)=0 $, we can rewrite the Josephson energy as
\begin{equation}\label{key}
H_J=-E_J(\phi) \cos\hat{\varphi}.
\end{equation}
We approximately regard $ e^{i\hat{\phi}} $ as the phase operator which adds an electron to the topological bulk segments, i.e. $  e^{i\hat{\varphi}}\ket{N_S}=\ket{N_S+1} $.
For the coupling quantum dot, the effective Hamiltonians are \cite{PhysRevB.95.235305}
\begin{eqnarray}\label{key}
H_{\text{QD}}^{\text{eff}}&=&h\hat{n_f}+\varepsilon_C(\hat{n_f}-n_g)^2,\quad \hat{n_f}=f^\dagger f,\\
\label{2MZM-qd-tunnel}
H_{\text{t,QD}}&=&-i\frac{e^{-i\varphi/2}}{2}(t_1f^\dagger \gamma_1+t_2 f^\dagger \gamma_{2N})+\text{h.c}.
\end{eqnarray}
where $ f^\dagger $ is the creation operators of the quantum dot.
In order to find the relationship between overall induced charge $ n_g $ of the quantum dot and the energy of the whole structure, we will diagonalize $H_{\text{tot}}=H_{C}+H_{J}+H_{\text{QD}}^{\text{eff}}+H_{\text{t,QD}}$ numerically, as sketched in Fig.~\ref{fig:ngvse}. Similar method has been used in Ref.~\cite{PhysRevB.95.235305}, while we consider the alterable Josephson energy here.

\subsection{\label{sec:majorana} Toric code and thermal baths}
\begin{figure}[htbp]
	\centering
	\includegraphics[width=0.75\textwidth]{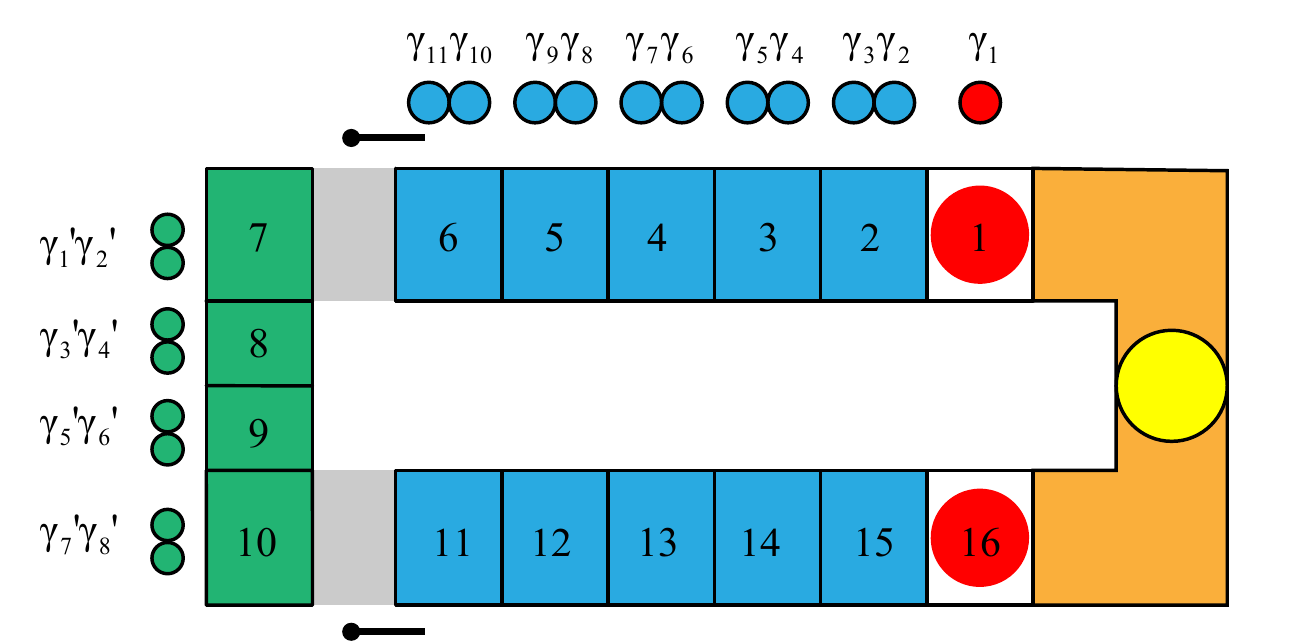}
	\caption{Box representation of an island contains two MZMs. Blue box is the fermion mode for quasi-particle $ d_j $ in topological bulk segment, green box is the fermion mode for electron in nontopological backbone, red dot box is the Majorana zero mode at the edge, yellow dot is the quantum dot coupled to the island. In this box model, the number on the boxes labels the site of the quasi-particles or the fermion mode. Here we use the amount of boxes to denote the length of the island and the backbone, i.e. $L=16$ and $ L_{\text{b}}=4 $.}
	\label{fig:structure1gamma}
\end{figure}

To discuss the lifetime of Majorana code in the island, we firstly build up the box representation model, which follows the method in Ref.~\cite{PhysRevB.92.115441}. As sketched in Fig.~\ref{fig:structure1gamma}, the boxes (blue) in the topological bulk segments represent the state of quasi-particles whose annihilation and creation operators are $ d_i $ and $ d_i^\dagger $ respectively. The right edge boxes (red circle inside) represent the unpaired MZMs at the edge of the topological bulk segments. The boxes in backbone (green) are the normal electron fermionic modes. The box representation is straightforward to encode the system with Kitaev's toric code. Boxes in the topological bulk segments and nontopological backbones are the stabilizer of the island (except the MZMs boxes) , thus measurement of these boxes will not change the state of the system. However, the boxes at the edge containing MZMs are not the stabilizers and the states are invisible.

The two degenerate ground states of the island sketched in Fig.~\ref{fig:structure1gamma} are $ \ket{0}=\ket{p_{12}=-1} $ and $ \ket{1}=\ket{p_{12}=1} $, where $ p_{12} $ is the eigenvalue of $ i\gamma_{1}\gamma_{2} $. The stabilizer operators $ S^{\text{top}}_i=i\gamma_{2i}\gamma_{2i+1}=2d_{i}^\dagger d_{i}-1 $ are used to defined the states ($ \ket{0_i} $ or $ \ket{1_i} $) of the box in the topological bulk segment at site $ i $, and the notation $ \psi $ is used to denote the excited state $ \ket{1_i}=d_i^\dagger \ket{0_i} $ in the box \cite{PhysRevB.92.115441}. For the nontopological backbone, stabilizer operators take the form $ S^{\text{nontop}}_i=i\gamma_{2i-1}\gamma_{2i}=c_i^\dagger c_i-1/2 $ and the notation $ \psi' $ is used to denote the excited state $ \ket{1_i'}=c_i^\dagger\ket{0_i'} $. The ground-state subspace of the stabilizer operators in the bulk segments and backbone forms a stabilizer code, which can be regarded as 1D Kitaev toric code.

The Hamiltonian of an open quantum system coupling to thermal baths can be written as: \cite{breuer2002theory}
\begin{equation}\label{key}
H=H_\text{S}+H_\text{B}+H_\text{SB}
\end{equation}
where $ H_\text{S} $ and $ H_\text{B} $ are the Hamiltonians of the system and the thermal baths respectively. $ H_\text{SB} $ denotes the interaction between the system and the thermal baths. The local operators of the system are distinct in different parts of the boxes sketched in Fig.~\ref{fig:structure1gamma}. In this paper, we introduce two external thermal baths, i.e. the non-interacting bosonic bath and fermionic bath. We use $ B_i^\dagger $ ($C_i^\dagger$) and $ B_i $ ($ C_i $) to denote the creation and annihilation operators of the bosonic (fermionic) bath. Both baths are local and can be respectively written as:
\begin{eqnarray}\label{key}
H_\text{B}^{(1)}=\sum_{i}\omega_iB_i^\dagger B_i.\\
H_\text{B}^{(2)}=\sum_{i}\varepsilon_iC_i^\dagger C_i.
\end{eqnarray}
Following the model in \cite{PhysRevB.92.115441}, the interaction between bosonic thermal bath and the island is assumed to couple to effective charge of the island:
\begin{eqnarray}\label{Interaction1}
H_{\text{SB}}^{(1)}=-\sum_i B_i\otimes (2c_i^\dagger c_i-1)
\end{eqnarray}
 As the Hamiltonian of the topological bulk segments can be diagonalized by the quasi-particle operators $ d_i^\dagger $ and $ d_i $, the bosonic bath leads to the $ \psi $ excitation creation, annihilation in pairs or shifting in the topological bulk segments, while it do not work on the nontopological backbone \cite{PhysRevB.92.115441}.

The interaction of fermionic thermal bath and the island is assumed to be
\begin{equation}\label{Interaction2}
H_{\text{SB}}^{(2)}=\sum_{i}t_iC_i^\dagger\otimes c_i+\text{h.c.}
\end{equation}
where $ t_i $ is the tunneling amplitude. The charging energies of topological bulk segments leads to large Coulomb blockade (i.e. $ t_i $ is small) \cite{PhysRevB.95.235305}, thus the interaction between the fermionic thermal bath and the topological bulk segments is neglected in this paper. The nontopological backbone is susceptible to the fermionic thermal bath which will induce the $ \psi' $ excitation creation and annihilation individually.

The interaction bosonic and the fermionic baths on the quantum dot is quite complicated \cite{de2016thermoelectric}. For simplicity, we focus on the effect of the fermionic bath, which can be written as
\begin{equation}\label{key}
H_{\text{SB}}^{(3)}=t_{d}C^\dagger\otimes f+\text{h.c.}
\end{equation}
where $ t_d $ is the tunneling amplitude between the the fermionic bath and the quantum dot.

\begin{figure}[htbp]
	\centering
	\includegraphics[width=0.95\textwidth]{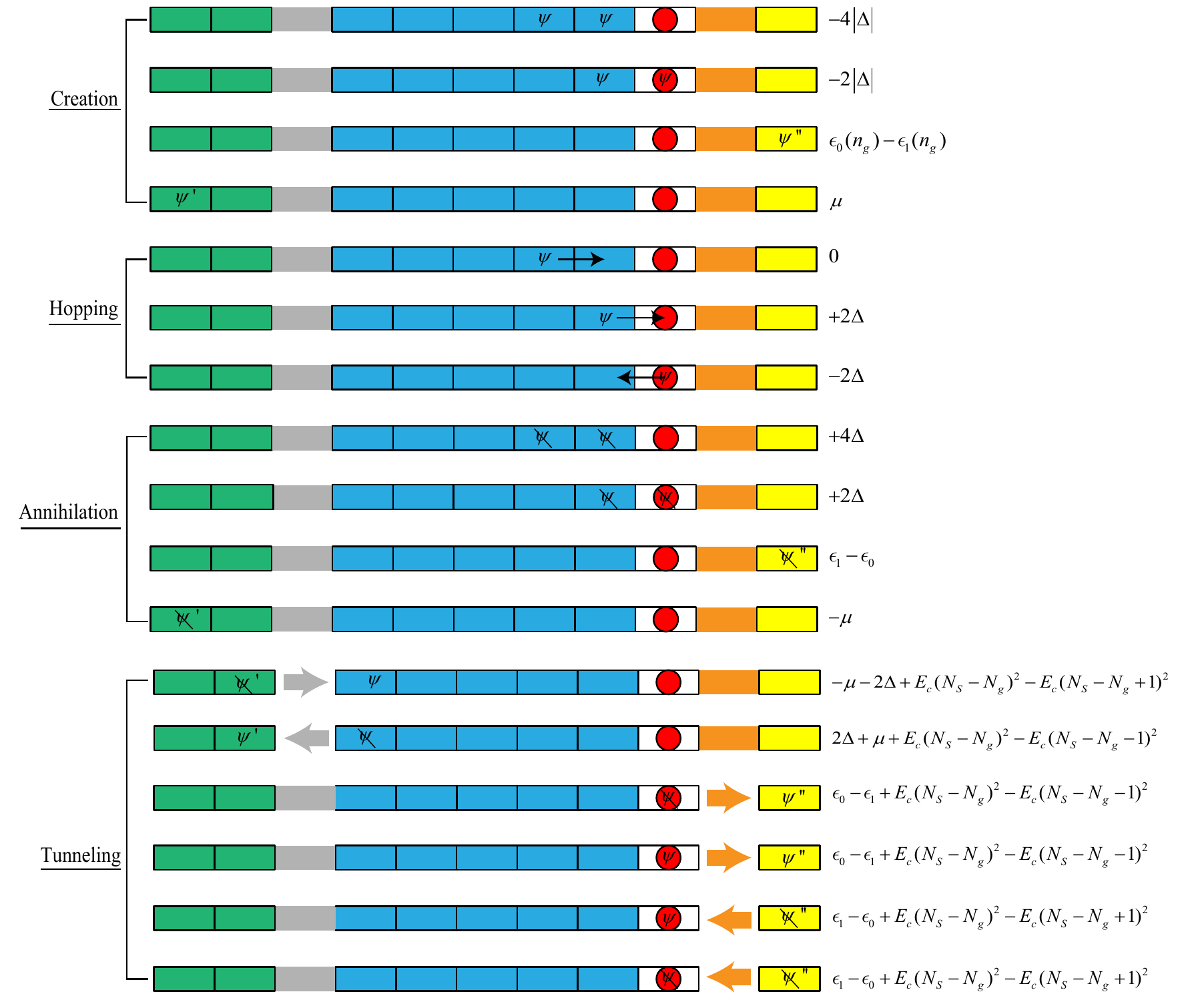}
\caption{Besides the cases of hooping, creation and annihilation considered in Ref.~\cite{PhysRevB.92.115441}, the parity-breaking interaction between the thermal baths and the non-topological backbone has been taken into consideration here, as well as the tunneling processes. Arrows in the tunneling block means that the charge is hooping with the fermionic tunneling.}
	\label{fig:omegadifferentcases}
\end{figure}

Assuming that the tunneling Hamiltonian between the backbone and the bulk segment through the valves are affective by the strong
hybridization on the left side, thus it takes the standard form
\begin{equation}\label{key}
H_{\text{t}}=-\Gamma (c^\dagger d_\text{end}+\text{h.c.}),
\end{equation}
where $ c^\dagger $ is the electron creation operators at the edge of the nontopological backbone and $ d_\text{end} $ is the adjacent quasi-particle annihilation operator of the topological bulk segment. Noting that
In this model, we have assumed that all the tunneling processes are related to the two kinds of thermal baths, which will be taken into consideration in Monte Carlo simulation.
One can describe an open system in thermal baths by Pauli master equation \cite{PhysRevB.92.115441,breuer2002theory}
\begin{equation}\label{pauli master equation}
\frac{\rm d}{\rm dt}P(n,t)=\sum_m [W(n|m)P(m,t)-W(m|n)P(n,t)].
\end{equation}
Here $ W(m|n)=\gamma(\omega)|\bra{m} A(\omega)\ket{n}|^2 $ is the error transition rate, and we consider the Ohmic bath correlation functions
\begin{equation}\label{key}
\gamma(\omega)=\kappa|\frac{\omega}{1-exp(-\beta\omega)}|,
\end{equation}
where $\omega=E_n-E_m$ denotes the energy difference between state $ \ket{n} $ and $ \ket{m} $. In Fig.~\ref{fig:omegadifferentcases}, we list $ \omega $ of different interaction cases.
\section{\label{sec:errorcorrection}Error correction and parity correction}
The string operator could create the excitations $ S_j=-1 $ and $ S_k=-1 $ on sites $ j $ and $ k $ respectively. As the positions of $ \psi $ or $ \psi' $ in bulk segments and backbone are detectable as the error syndromes, string operators can be applied to fuse the excitations on thses sites, which takes the form \cite{kitaev2003fault}
\begin{equation}\label{key}
S_{j,k}=\gamma_{2j+1}\gamma_{2j+2}...\gamma_{2k}.
\end{equation}

 Hard-decision renormalization group (HDRG) decoders have been used in topological error correction \cite{PhysRevA.93.022318}. In this paper, a simple form of HDRG decoder in Ref.~\cite{wootton2015simple} is implemented. Similar method has been studied in \cite{PhysRevB.92.115441} and \cite{PhysRevLett.115.120402}, while we consider more complex cases here.

\begin{figure}[htbp]
	\centering
	\includegraphics[width=0.95\textwidth]{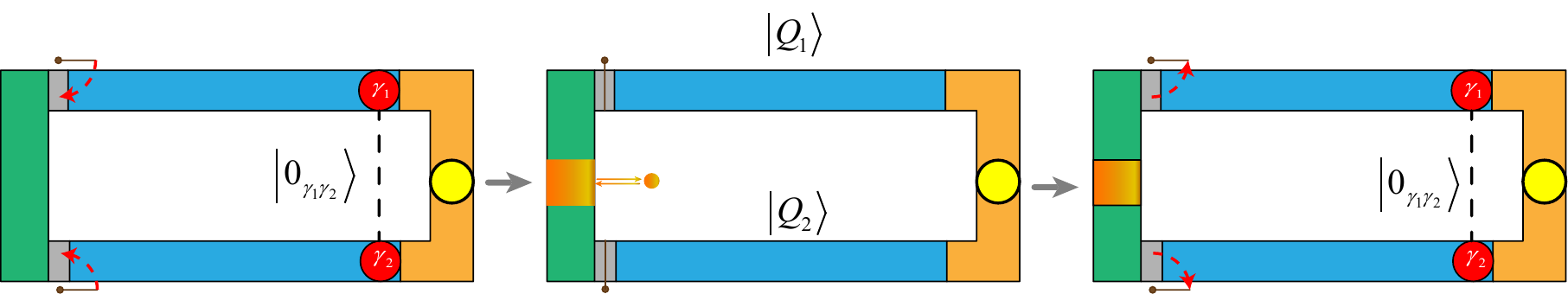}
	\caption{Protocol to correct the parity of the island. Firstly, the valves are closed and the excitation parity is turned into detectable charge parity states. Secondly, if the charge parity states are different, a weak photon pulse is injected into the mid area of the backbone which would flip the excitation parity of the island. Finally, the valves are closed and the bulk segments revert to be topological.}
	\label{fig:structure1paritycorrection}
\end{figure}

The errors caused by local thermal baths interaction in the bulk topological segments (blue boxes) are assumed to be parity conserving, while the interactions on backbone and the electron tunneling from the quantum dots will break the parity of the island. When the parity of the island is flipped, the HDRG decoder will draw the error excitations to the MZMs of the island and cause the error topological states. Thus we need to correct the parity firstly. As depicted in Fig.~\ref{fig:structure1paritycorrection}, it contains the following procedures:

(1) Initialize the island with the open valves. The charging energy of bulks is quenched and the bulks are in topological states with MZMs $ \gamma_{1} $ and $ \gamma_2 $.

(2) Close the valves. The strong coulomb effect would draw the system into nondegenerate charge states $ \ket{Q_1} $ and $ \ket{Q_2} $. Odd number of electron tunneling would flip the charge parity, i.e, $ \ket{Q_1,P_1} \rightarrow \ket{Q_1 \pm (2N+1),-P_1}$, while the even number of electron tunneling would not, i.e, $ \ket{Q_1,P_1} \rightarrow \ket{Q_1 \pm (2N),P_1}$. With the use of the quantum dot on the right side, charge sensors is implemented and the charge parities of $ \ket{Q_1} $ and $ \ket{Q_2} $ can be detected, as the experiment proposal in Ref.~\cite{RN62}. If the parities are different, a weak photon pulse is injected into the middle region of the backbone. The energy of the photon pulse is precisely controlled so that only one electron excitation or annihilation would happen in the backbone to flip the parity of the island.

(3) Open the valves and the bulk segments recover back to the topological state with MZMs at the edge.

\begin{figure}[htbp]
	\centering
	\includegraphics[width=0.75\textwidth]{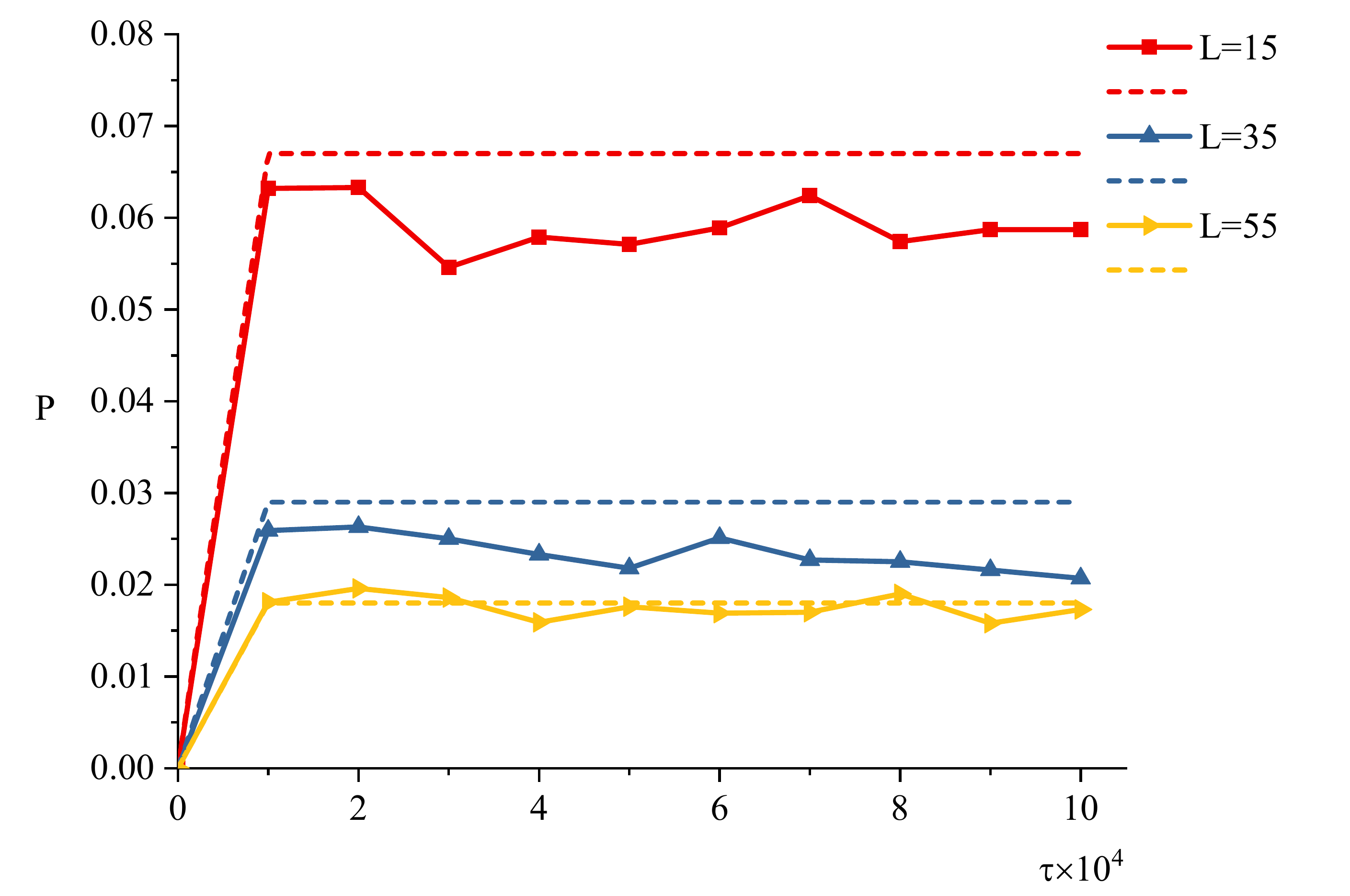}	
    \caption{Probability of the logical X or Y error as a function of time $ \tau $. Here we compare the infection of different island length $ L=15,35,55 $, while the length of the backbone is fixed by $ L_b=4 $. (a) We take $ \mu=-2 $ and $ W=|\mu| L_{\text{b}}=8 $ in high-temperature regime. (b)We simulate the 2-MZM island with the Monte Carlo method below. We take $ \Delta=0.7 $ and $ \mu=-2 $ to draw the error transition rates approximate in the backbone and the bulk segment (make it similar with the fixed error transition rate $ W $ in the analytical results). The other parameter values are listed in Table \ref{2-MZMislandParameters}. }
    \label{fig:pxy}
\end{figure}

It is important to note that this parity correction protocol can not correct the parity-breaking errors aroused by the backbone-thermal baths interaction, because $ \ket{Q_1,P_1}$ and $ \ket{Q_2,P_2}$ will not change when any creation or annihilation is happened in the backbone, i.e, it is undetectable for this protocol. We will see that the parity breaking in backbone would still induce the logical X or Y error. We can use the Jordan-Wigner transformation to estimate the probabilities of logical X or Y errors. It is convenient to rewrite the nontopological backbone Hamiltonian in terms of spin operators
\begin{equation}\label{key}
H_\text{b}=-\frac{i}{2}\sum_{i=1}^{N}\mu\gamma_{2i-1}\gamma_{2i}=\frac{\mu}{2}\sum_{i=1}^{N}\sigma^z_i,
\end{equation}
where we have used the relationships
\begin{eqnarray}\label{key}
\gamma_{2i-1}=(\prod_{k=1}^{i-1}\sigma_k^z)\sigma_i^x,\\
\gamma_{2i}=(\prod_{k=1}^{i-1}\sigma_k^z)\sigma_i^y.
\end{eqnarray}
The number of the spin flips in the backbone follows Poisson distribution with mean $ N=Wt $, where $ W $ is the total rate of error transitions caused by the interaction between the backbone and thermal baths. For simplicity, we use $ L_{\text{b}} $ to denote the length of the backbone, and we take $ W=|\mu| L_{\text{b}} $ in high-temperature regime. We can obtain the probability of number $ k $ of the spin flips in the backbone.
\begin{equation}\label{key}
P(k)=\frac{(Wt)^k}{k!}e^{-Wt}.
\end{equation}
An odd number of spin flips in the non-topological backbone will induce the logical X or Y errors, thus we can obtain
\begin{equation}\label{key}
P_X=P_Y=\frac{1}{2}\frac{L_{\text{b}}}{L}\sum^{\infty}_{k=odd}\frac{(Wt)^k}{k!}e^{-Wt}=\frac{L_{\text{b}}}{4L}(1-e^{-2Wt})
\end{equation}
where $ L $ is the total length of the island. This expression shows that the value of $ P_X $ or $ P_Y $ tends to a fixed value, which is expected to be $ L_b/4L $ when $ \tau $ is large enough, as sketched in Fig.~\ref{fig:pxy}. It will be verified in Monte Carlo simulation. $ P_X $ or $ P_Y $ is not only related to the length of the backbone but also relevant to the chemical potential $ \mu $ and the pairing potential $ \Delta $.
\section{\label{sec:montecarlo} Monte Carlo simulation}
To study the property of the error correction in scalable island, We use the Pauli master equation (\ref{pauli master equation}) to describe both of the islands and simulate the 2-MZM and 4-MZM modified islands with standard Monte Carlo method. As the parameter values of $ \Delta $ and $ \mu $ would act as the key role of the Monte Carlo results, we firstly discuss the probability of correctness as a function of these parameters. Then we chose the reasonable parameter values to study the 2-MZM modified island. Finally, as the 4-MZM modified island is coupled to two quantum dots, more electron tunneling situations will be taken into consideration.
\subsection{Two-MZM island simulation}
The Monte Carlo study of the scalable design will begin with the simplest structure as we sketch it in boxes form in Fig \ref{fig:structure1gamma}. We use the standard residence time Monte Carlo algorithm to study the Pauli master equation \cite{doi:10.1002/9783527683147.ch9}.
The simulation is implemented according to the following steps:

(1). Initialize the relevant parameters of the system.
\begin{table}
	\captionof{table}{\centering Parameters in Monte Carlo simulation of 2-MZM modified island}
	\begin{ruledtabular}
		\begin{tabular}{lcdrlcdr}
		
		 $ \beta $  &  $ N_g $ &  E_C &  $h$ &  $ \varepsilon_C $ & $n_{g}$ &  t_{1,2}  & $ \Gamma $ \\	
		 	\colrule
		 2 & 0.1 & 1 & 0.5 & 5 & 0.35 & 0.8 & 0.8\\
\end{tabular}
\end{ruledtabular}
\label{2-MZMislandParameters}
\end{table}

(2). Determine the time $ \delta\tau=-\ln(u)/W_{\text{tot}} $ for the next jump, where $ u $ is a random number distributed in the interval $ (0,1) $ uniformly. $ W_{\text{tot}}=\sum_m W(m|n) $ is the total error transition rate caused by the interaction between the system and the thermal baths for a system state $ \ket{n} $.

(3). Update the simulation time to $ \tau+\delta \tau $ and if $ \tau+\delta \tau\leq \tau_{\rm sim} $, go to step $ 4 $, or go directly to step $ 5 $.

(4). Implement the error transition randomly on the system according to the probability. Go back to step $ 2 $.

(5). Detect the stabilizer operators and apply the error correction protocol including the parity correction and the HDRG algorithm. Record the state of the system. In order to simulate the excitation caused by the weak photon pulse injection, the parity correction here is implemented through flipping one green box of backbone randomly if the charge parity is different.

(6). Repeat step (1) to (5) thousand of times and record the rates of correctness and different errors.
\begin{figure}[!htbp]
	\centering
 	\includegraphics[width=0.75\textwidth]{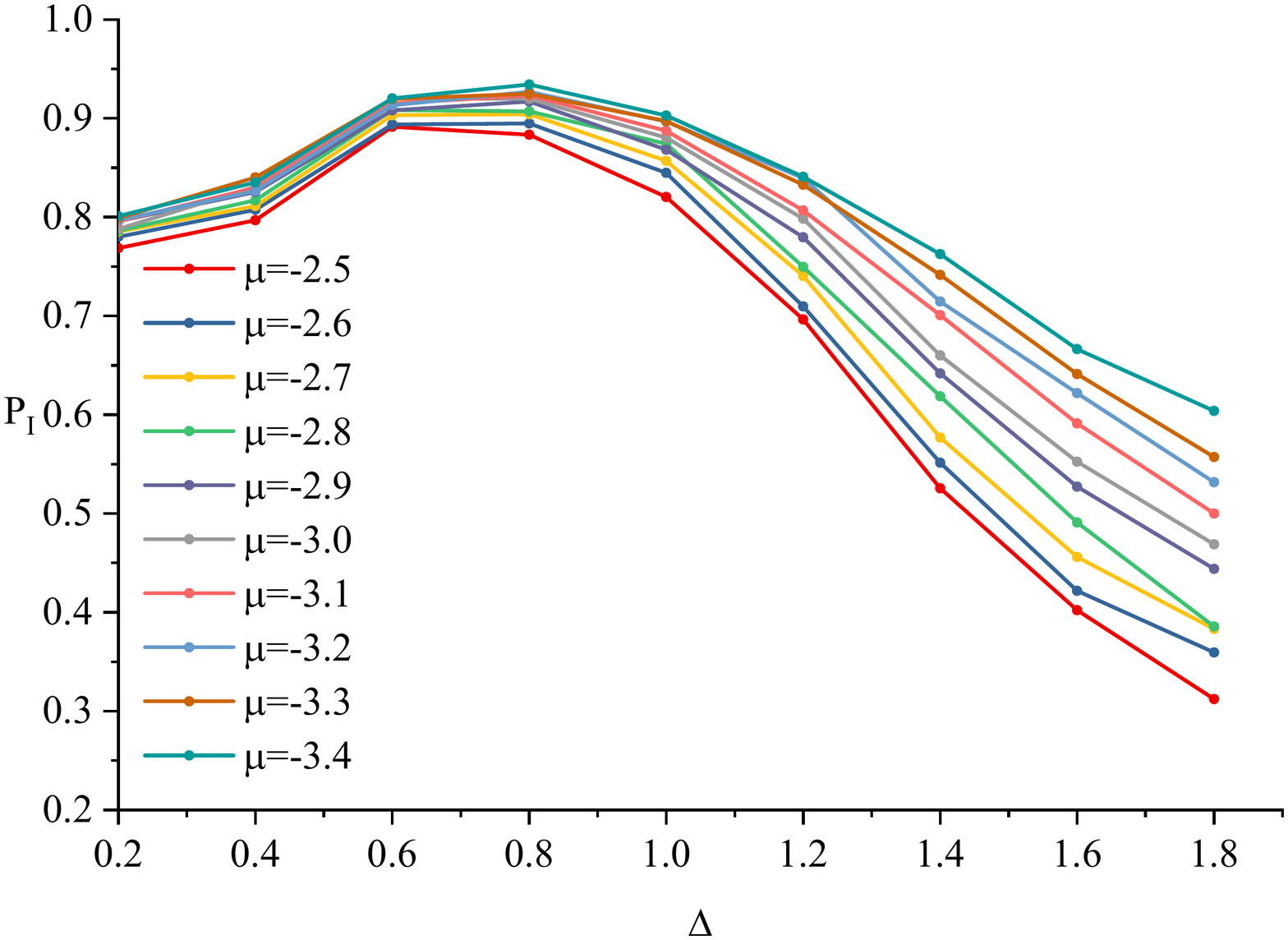}
	\caption{Monte Carlo result of the probability of correctness $ P_I $ of the modified 2-MZM island with $ L=108 $. We set $ \tau_{\text{sim}}=10^5 $. We take $ \Delta $ within the range of $ 0.5 $ to $ 2.0 $ and $ \mu $ within the range of $ -2.5 $ to $ -3.4 $. }
	\label{fig:pimudelta}
\end{figure}

In Fig \ref{fig:pimudelta}, we apply the Monte Carlo method on the 2-MZM island and change the parameter $ \Delta $ within the range of $ 0.2 $ to $ 1.8 $ and $ \mu $ within the range of $ -2.5 $ to $ -3.4 $. For fixed $ \mu $, the value of $ P_I $ increases with $ \Delta $ firstly, and then decreases. It implies that the value of $ \Delta $ could be fine tuned to obtain the maximal probability of correctness in experience. The logical X and Y errors are suppressed by the decreasing of $ \mu $, thus for the fixed $ \Delta $, $ P_I $ increases with the decreasing of $ \mu $.

\begin{figure}[htbp]
	\centering
	\includegraphics[width=0.75\textwidth]{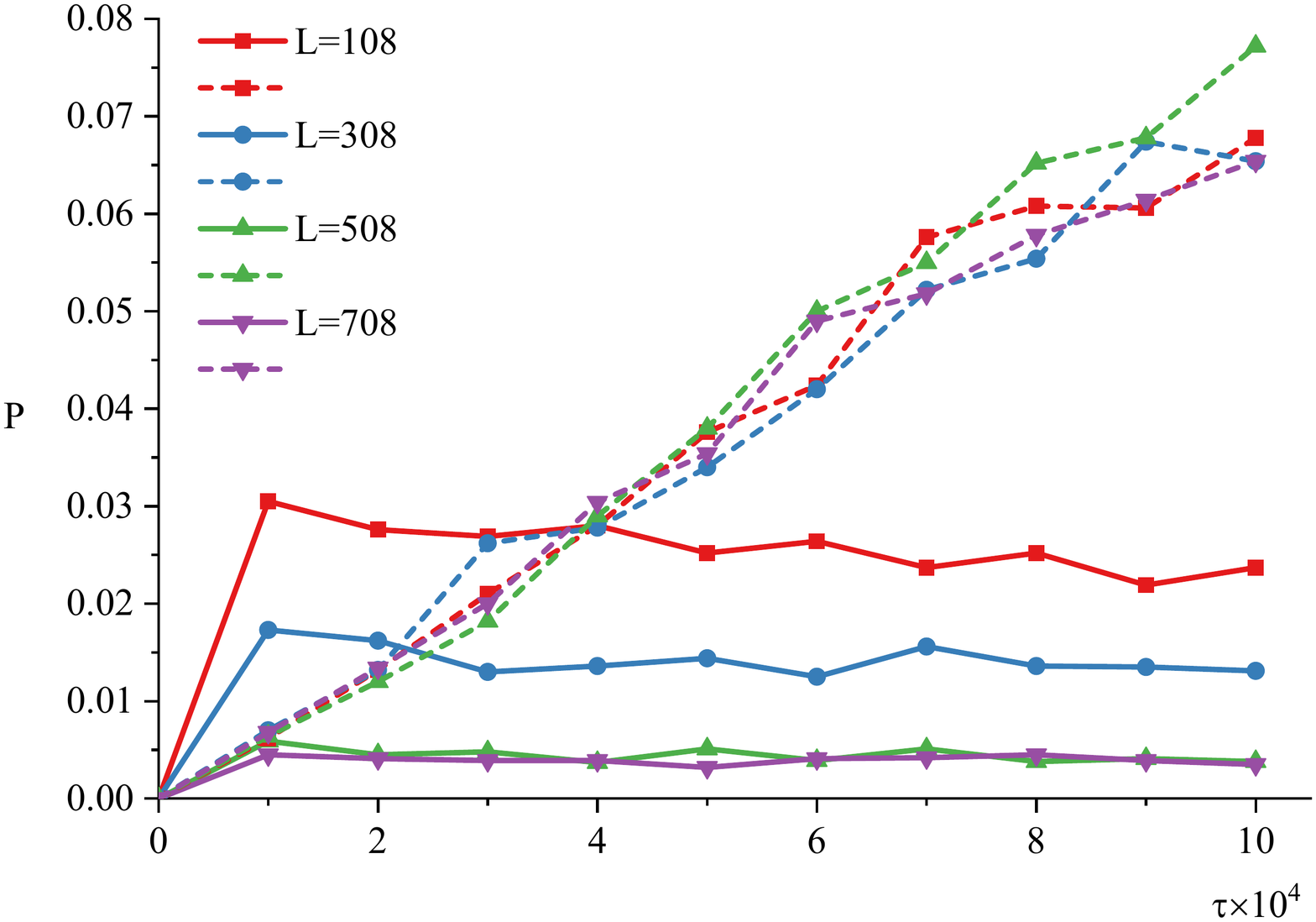}
	\caption{The error probabilities $ (P_X+P_Y)/2 $ (solid lines) and $ P_Z $ (dashed lines) as functions of $ \tau $ of the 2-MZM modified island with $L=108, 308,508, 708$ coupled to a quantum dot. We use $ \Delta=1 $ and $ \mu=-3 $ for the system. The length of backbones are fixed with 8 boxes. We set the simulation time $ \tau_{\rm sim}=10^5 $ here.}
	\label{fig:structure1montecarloresult}	
\end{figure}

We use $ \Delta=1 $ and $ \mu=-3 $, and simulate the island with $ L=108,308,508, 708 $ to observe the influence of the island length. In Fig~.\ref{fig:structure1montecarloresult}, we exhibit that the probabilities of logical errors with different length as functions of $ \tau $. We use $(P_{X}+P_{y})/2$ to describe these errors here. It can be observed that the probability of logical X and Y errors decreases with the length of the island and it would meet a threshold, which corresponds to our discussion in Sec.~\ref{sec:errorcorrection}. The parity breaking excitations in the bulk segments are offset by our parity correction method and only the odd number of anyon excitations in the backbone would cause the logical X or Y errors. The probabilities of logical X and Y errors are similar because the generation of the excitations in backbone is random and the HDRG decoder would draw excitations to the upper MZM or the lower MZM randomly.   Furthermore, the Monte Carlo results show that there is a threshold (about $0.005$) for $ (P_{X}+P_{y})/2$, as we lengthen the 2-MZM modified island. The probability of the logical Z error has no significant change with the length of the island because the anyon tunneling is the major source of the logical Z error, which is not apparently affected by the length of the island.
\subsection{\label{sec:fourMZMisland} Four-MZM island simulation}
\begin{figure}[htbp]
	\centering
	\includegraphics[width=0.75\textwidth]{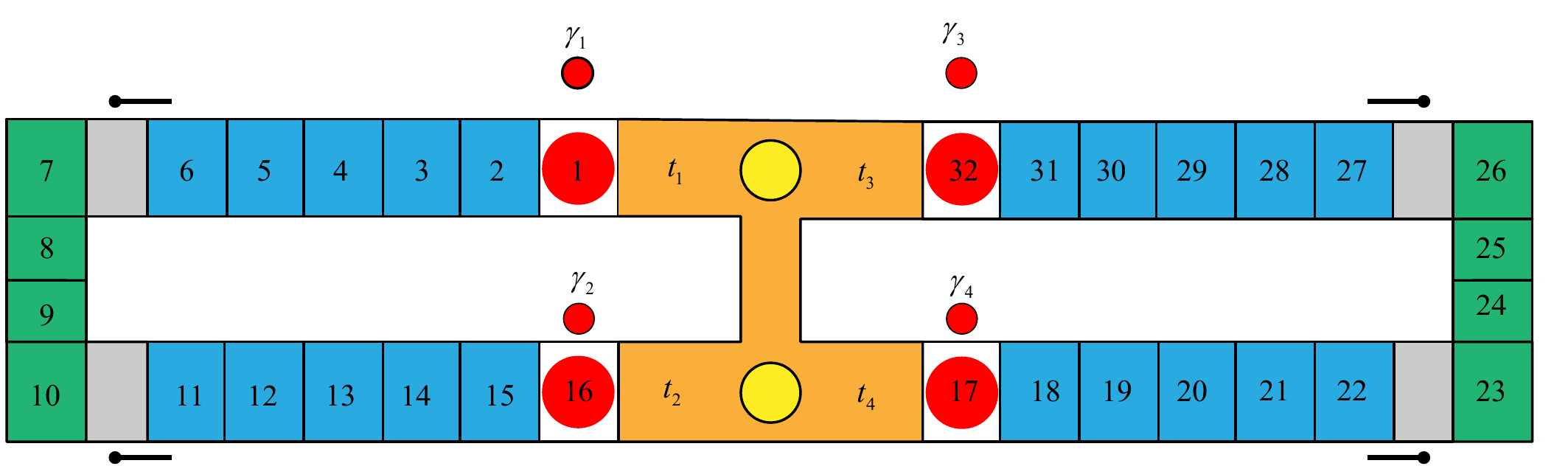}
	\caption{Box representation of the four-MZM modified island with $ L=32 $ and $ L_{\text{b}}=4 $. The four-MZM parity is determined by $ p=(i\gamma_{1}\gamma_{2})(i\gamma_{3}\gamma_{4}) $. The upper quantum dot is tunnel coupled to $ \gamma_{1} $ and $ \gamma_{3} $ with the tunneling amplitude $ t_1 $ and $ t_3 $ respectively. And the lower quantum dot is tunnel coupled to $ \gamma_2 $ and $ \gamma_4 $ with $ t_2 $ and $ t_4 $. The mutual charging energy of the two quantum dots is taken into consideration \cite{kitaev2001unpaired}. Therefore, when we calculate the probability of the interaction between the quantum dots and the thermal baths, the occupations of these two quantum dots are taken into account.}
	\label{fig:structure2}
\end{figure}

Similar discussion is implemented on the configuration of the modified 4-MZM island in Fig.~\ref{fig:structure2}. In this configuration, 4 MZMs and the mutual interaction between the two quantum dots are taken into consideration. Thus the effective Hamiltonians for the quantum dot and tunneling are \cite{PhysRevB.95.235305}
\begin{eqnarray}\label{key}
H_{\text{QD}}^{\text{eff}}&=&\sum_{i=1,2}h_i\hat{n_{f,i}}+\varepsilon_{C,i}(\hat{n_{f,i}}-n_{g,i})^2+H_{\text{m}}\\
H_{\text{t,QD}}&=&-i\frac{e^{-i\phi_1/2}}{2}(t_1 f_1^\dagger \gamma_1+t_2 f_2^\dagger\gamma_2)-i\frac{e^{-i\phi_2/2}}{2}(t_3 f_1^\dagger \gamma_3+t_2 f_2^\dagger\gamma_4)+\text{H.c.},
\end{eqnarray}
 where $ H_{\text{m}}=\epsilon_M(\hat{n}_{f,1}-n_{g,1})(\hat{n}_{f,2}-n_{g,2}) $ is the mutual interaction Hamiltonian of the two quantum dots. Electrons in each quantum dot have two directions to tunnel. Such kind of tunneling would break the parity of the island.

\begin{figure}[htbp]
	\centering
	\includegraphics[width=0.95\textwidth]{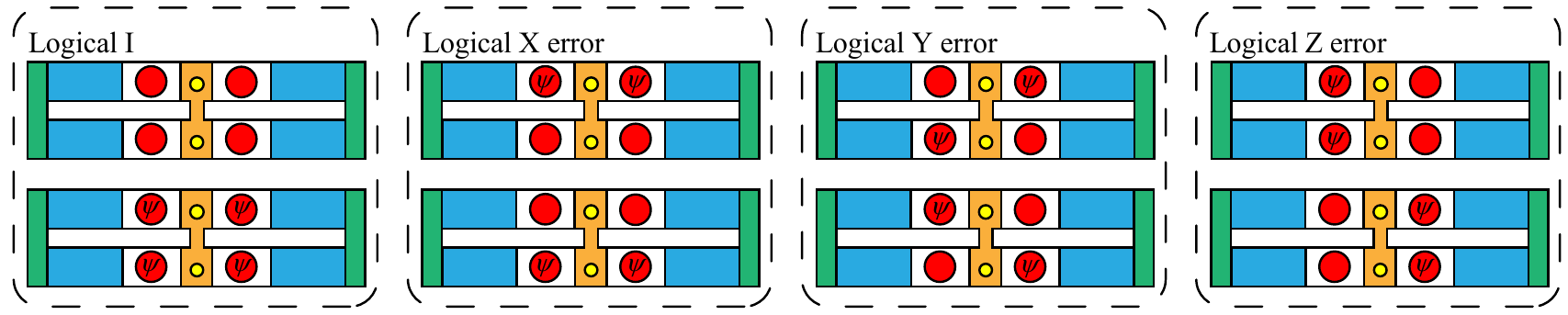}
	\caption{We represent Logical I (the correct cases), X, Y, and Z errors with the sites of excitations of MZMs. Every logical error corresponds to two excitation cases.}
	\label{fig:structure2logicalerror}
\end{figure}

The parity of 4 MZMs is determined by $ p=(i\gamma_{1}\gamma_{2})(i\gamma_{3}\gamma_{4}) $. For clarity, we use $ p=+1 $ and thus obtain $\ket{0}=\ket{0_{\gamma_{1}\gamma_{2}},0_{\gamma_{3}\gamma_{4}}} $ and  $\ket{1}=\ket{1_{\gamma_{1}\gamma_{2}},1_{\gamma_{3}\gamma_{4}}} $. We list all the logical errors in Fig.~\ref{fig:structure2logicalerror} and every logical error corresponds to two excitation cases. Besides the logical errors, any other excitation errors are classified as the else error.

The Monte Carlo simulation is implemented by the following steps:

(1). Initialize all parameters of the 4-MZMs modified island. The parameters values here are consistent with the ones of 2-MZM island.

\begin{table}[ht]
\centering
\caption{\centering Parameters in Monte Carlo simulation of 4-MZM modified island}

\begin{ruledtabular}
	\begin{tabular}{lcdrlcdrl}
		$ \beta $  &  $N_g$  &  E_C  & $ h_{1,2} $  & $ \varepsilon_{C,1,2} $ & $ n_{g,1,2} $ &  t_{1,2,3,4}  &$\Gamma$ & $\varepsilon_M$ \\
		\colrule
		 2 & 0.1 & 1 & 0.5 & 5 & 0.35 &0.8 & 0.8 & 0.5\\
\end{tabular}
\end{ruledtabular}
\label{4MZMparameter}
\end{table}

(2). Determine the time $ \delta\tau=-\ln(u)/W_{\text{tot}} $ for next jump.

(3). Update the simulation time to $ \tau+\delta \tau $ and implement the error transition randomly on the system according to the transition rate. If $ \tau+\delta \tau\leq \tau_{\rm sim} $, go to step (4) or go back to step (2).

(4). Detect the stabilizer operators and apply the parity correction proposal on both islands simultaneously, make sure that the valves are opened and closed at the same time. Then apply the HDRG algorithm to correct the error excitations.

(5). Repeat step (1) to (4) one thousand times and record the probabilities of correctness and different errors.

\begin{figure}[htbp]
	\centering
	\includegraphics[width=0.75\textwidth]{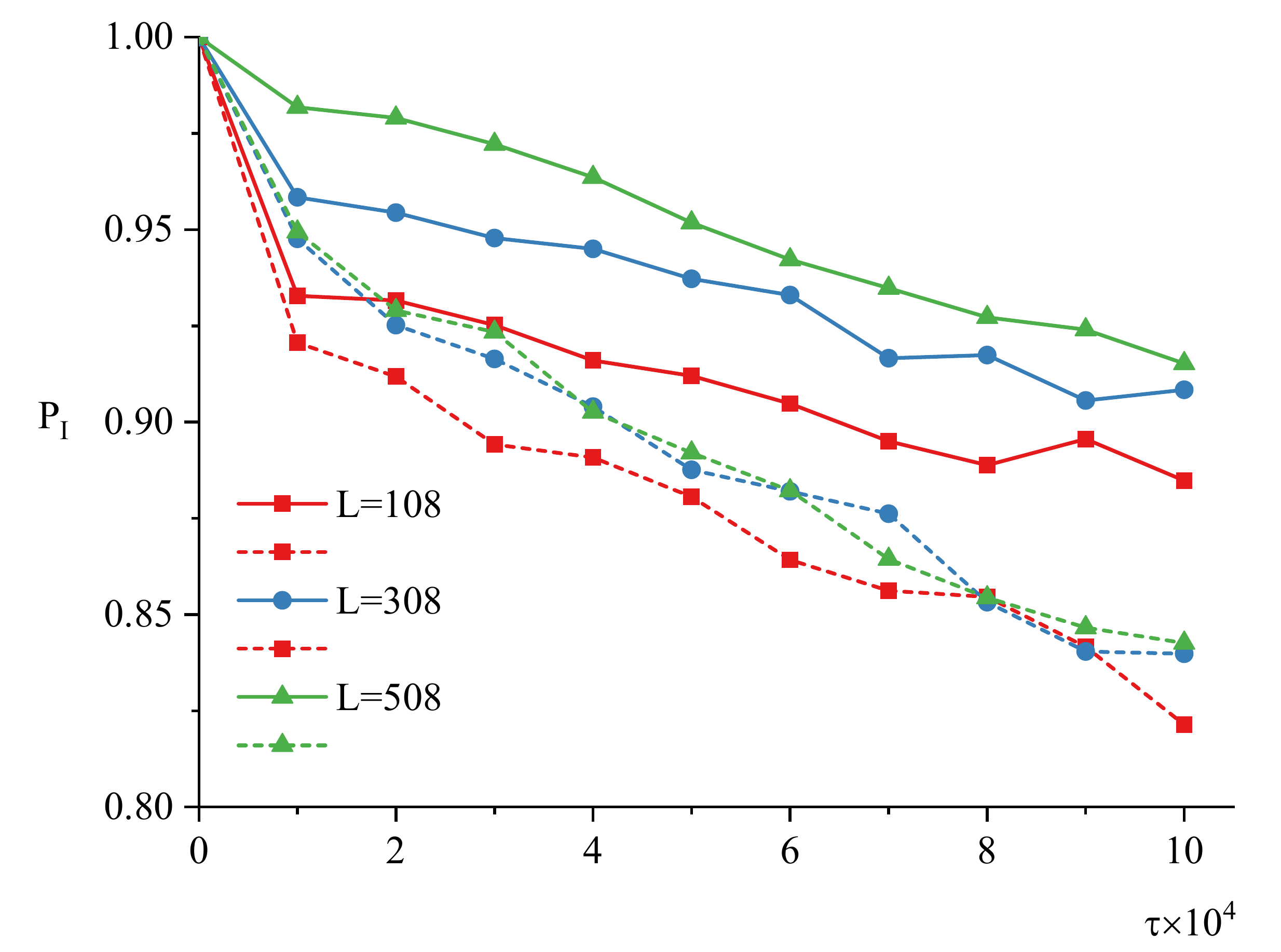}
	\caption{Probability of correctness ($ P_I $) of the 2-MZM and 4-MZM modified island with length $ L=108, 308, 508 $. The solid lines correspond to the 2-MZM modified islands while the dashed lines correspond to the 4-MZM ones. We use $ \Delta=1 $, $ \mu=-3 $ and other parameters values are listed in Table \ref{2-MZMislandParameters} and \ref{4MZMparameter}. The maximum time of simulation is set to be $ \tau=10^5 $.}
	\label{fig:pis1s2}
\end{figure}

In Fig \ref{fig:pis1s2}, we set $ L=108, 308, 508 $ to compare the probability of correctness ($P_I$) of the 4-MZM to the 2-MZM modified island. The length of nontopological backbones are set to 8 on both sides of the 4-MZM island. The dashed lines represent the Monte Carlo result of the 4-MZM modified island while the solid line curves for the 2-MZM ones which we have simulated in Fig \ref{fig:structure1montecarloresult}. The result shows that longer islands have higher probability $ P_I $ than the shorter ones. As we had discussed in Section \ref{sec:errorcorrection}, the probabilities of logical X and Y errors are suppressed by the length of the island. The result also shows that the 2-MZM modified island performs better than the 4-MZM one with the same length. That is because more tunneling channels and higher proportion of the nontopological backbones are contained in the the 4-MZM modified island.

\begin{figure}[htbp]
	\centering
	\includegraphics[width=0.75\textwidth]{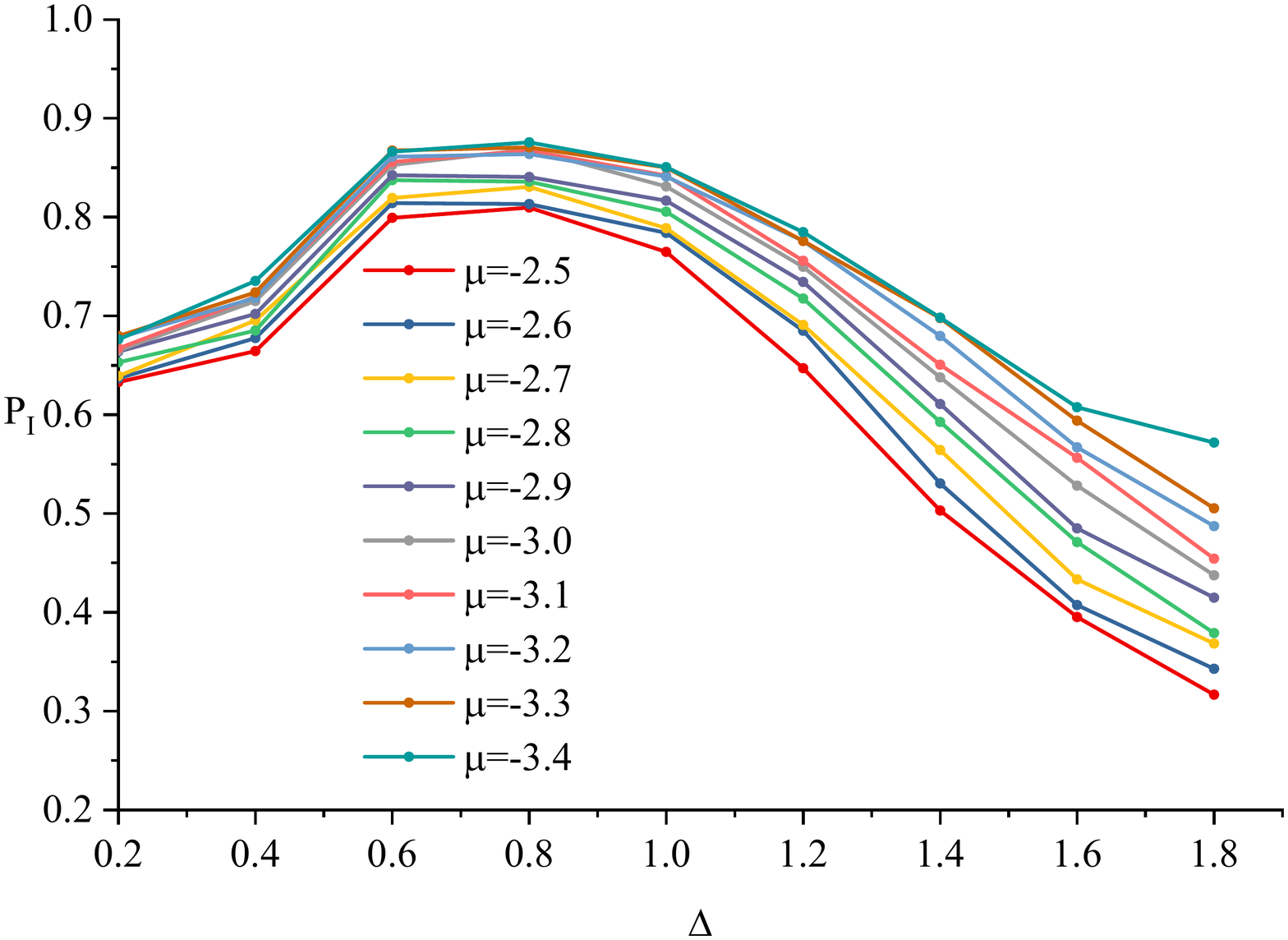}
	\caption{The Monte Carlo result of the 4-MZM modified island with different parameter values of $ \Delta $ and $ \mu $. The simulation time is $ \tau=10^5 $.  We take $ \Delta $ within the range of $ 0.5 $ to $ 2.0 $ and $ \mu $ within the range of $ -2.5 $ to $ -3.4 $, while the other parameter values are listed in Table \ref{4MZMparameter}.}
	\label{fig:s2pimudelta}
\end{figure}

Similar to the way of choosing parameter values in 2-MZM modified island, we change $\Delta $ within the range of $ 0.2 $ to $ 1.8 $ and $ \mu $ within the range of $ -2.5 $ to $ -3.4 $ in Fig~\ref{fig:s2pimudelta}. Compared to the 2-MZM modified island, the peaks of the curves are in the lower positions, which means that the 4-MZM ones needs smaller $ \mu $ for high probability of correctness. That is because the 4-MZM modified island is coupled to 2 quantum dots and has more tunneling channels. Proper $ \Delta $ and smaller $ \mu $ are beneficial to suppress the probability of the tunneling and the anyon excitations in the backbones.

\begin{figure}[htbp]
	\centering
	\includegraphics[width=0.75\textwidth]{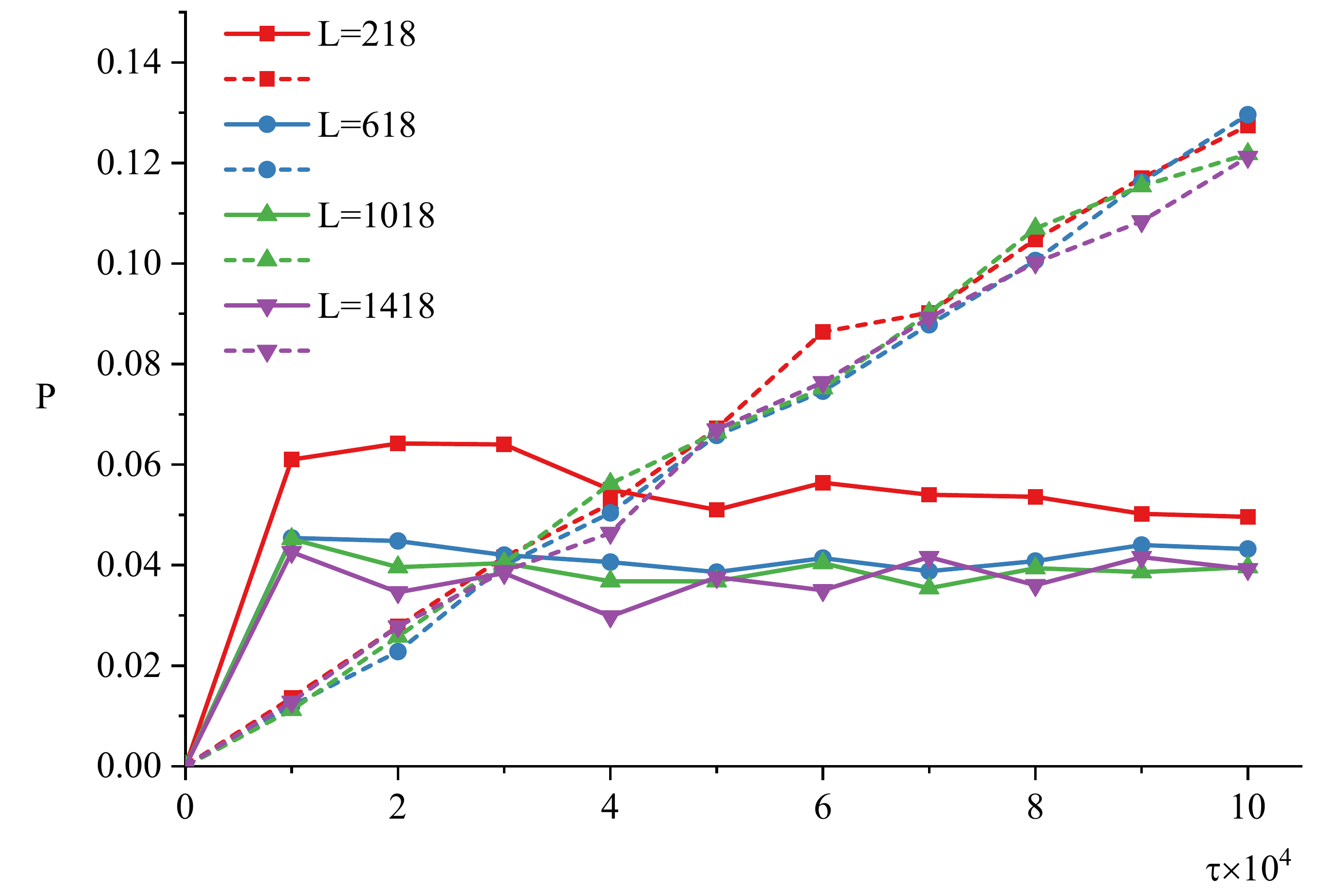}
	\caption{Probabilities of logical errors as functions of $ \tau $, which corresponds to the box model of the 4-MZM modified island with $L=216, 616, 1016, 1418$, and 2 quantum dots are taken into consideration here. We use $ \Delta=1 $, $ \mu=-3 $. The length of the backbones are $ L_b=8 $ on both sides, we change the length of the bulk segments while the length of backbones are remained unchanged. The solid lines represent the probabilities of logical Z gate error of the 4-MZM modified islands with different sizes.}	
	\label{fig:structure2montecarloresult}
\end{figure}
As we have discussed in the case of the 2-MZM modified island, though our parity correcting proposal can correct the QPPs in the bulk segments, it is still susceptible to the ones in the backbones, which are the primary source of else errors, as depicted in Fig.~\ref{fig:structure2montecarloresult}. Different from the 2-MZM cases, the major errors here are the logical Z gate and else errors.  Logical X and Y errors are almost suppressed entirely ($ \le10^-3 $), thus they are neglected. The probability of the else errors tends to a fixed value, which is suppressed by the length of the 4-MZM modified island. The threshold is around $ P_{else}\approx 0.035 $ with the increasing of the island length. The probabilities of the logical Z gate error are similar.

\section{\label{sec:conclusion} Conculsion}
In this work, we have modified the 2-MZM and 4-MZM islands in Ref.~\cite{PhysRevB.95.235305} for scalable quantum computing and proposed a parity correction scheme. Based on these modified islands, we have simulated the system in two local thermal baths with Monte Carlo method. The parity correction scheme is valid against the QPPs in the bulk segments. However, the parity breaking excitations in the backbones can still cause some annoying errors. These annoying errors can be suppressed by lengthening the topological bulk segments although it would meet a threshold. Furthermore, adjusting the chemical potential of the non-topological backbones and the pairing potential of the topological bulk segments is effective to reduce the probability of these errors.
The Monte Carlo results show that when our parity correction scheme is implemented, to a certain degree, larger size island has better error correction rate than the smaller one. However, if the the threshold of size is met, lengthening the island cannot improve the probability of correctness. Instead, it will lead to worse error correction rate because of the long-time application of the error correction algorithms, which is in agreement with the result of \cite{PhysRevB.92.115441, PhysRevA.93.022318, PhysRevX.4.031058, PhysRevX.4.011051}. Besides, small chemical potential $ \mu $ of the backbone and fine-tuned pairing potential $ \Delta $ of the bulk segment are required for high probability of correctness.

Gate-tunable valves and non-topological backbones are the essential ingredients of our proposal. With the help of these valves, the error excitation states are turned into the charge states, which are detectable in experiments. We take advantage of non-topological backbones to implement the anyons fusion by the string operators and topological fermion parity correction by the weak photon pulse. In this work we study a single island for simplicity. For large scale FTQC, several islands are needed to implement the braiding transformations through the measurement-only schemes. The lifetime of the Majorana based qubit is more complicated, and more error correction algorithms should be considered. However, the generalization would be straightforward. We expect that our modified island scheme could be applicable for the multiple islands error correction.
\begin{acknowledgments}
We would like to thank Zhongbo Yan for helpful discussions.
This work is supported by the National Natural Science Foundation of China (NSFC) under Grant No. 11875327, the Natural Science Foundation of Guangdong Province under Grant No. 2016A030313313, the Fundamental Research Funds for the Central Universities, and the Sun Yat-Sen University Science Foundation.
\end{acknowledgments}

\appendix
\section{Numeric diagonalization of $ H_{\text{tot}} $}
\begin{figure}[htbp]
	\begin{center}
		\begin{subfigure}{0.45\textwidth}
			\includegraphics[width=8.6 cm]{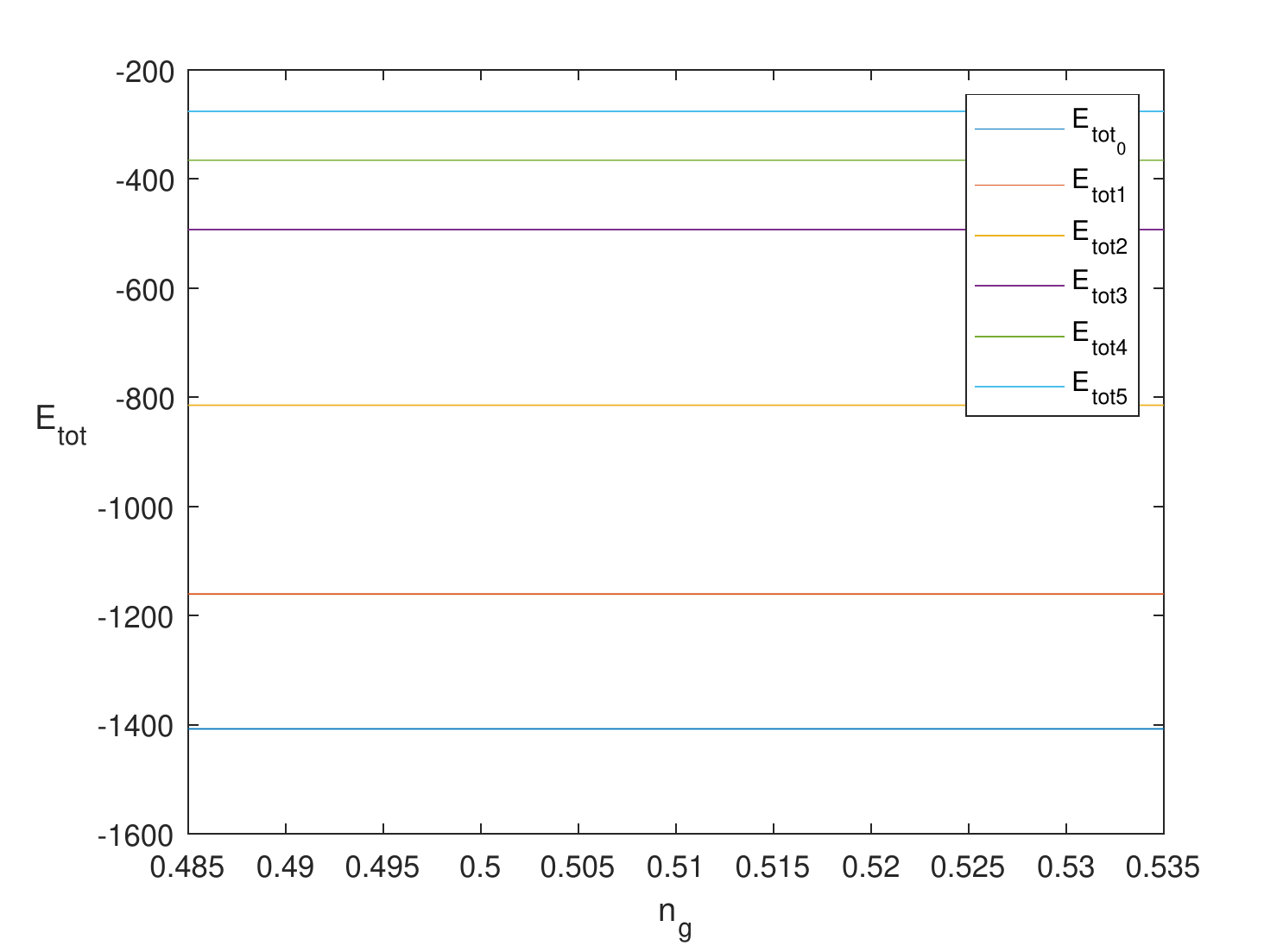}
			\caption{\centering $ E_J/ E_C=1000 $}
		\end{subfigure}
		\begin{subfigure}{0.45\textwidth}
			\includegraphics[width=8.6 cm]{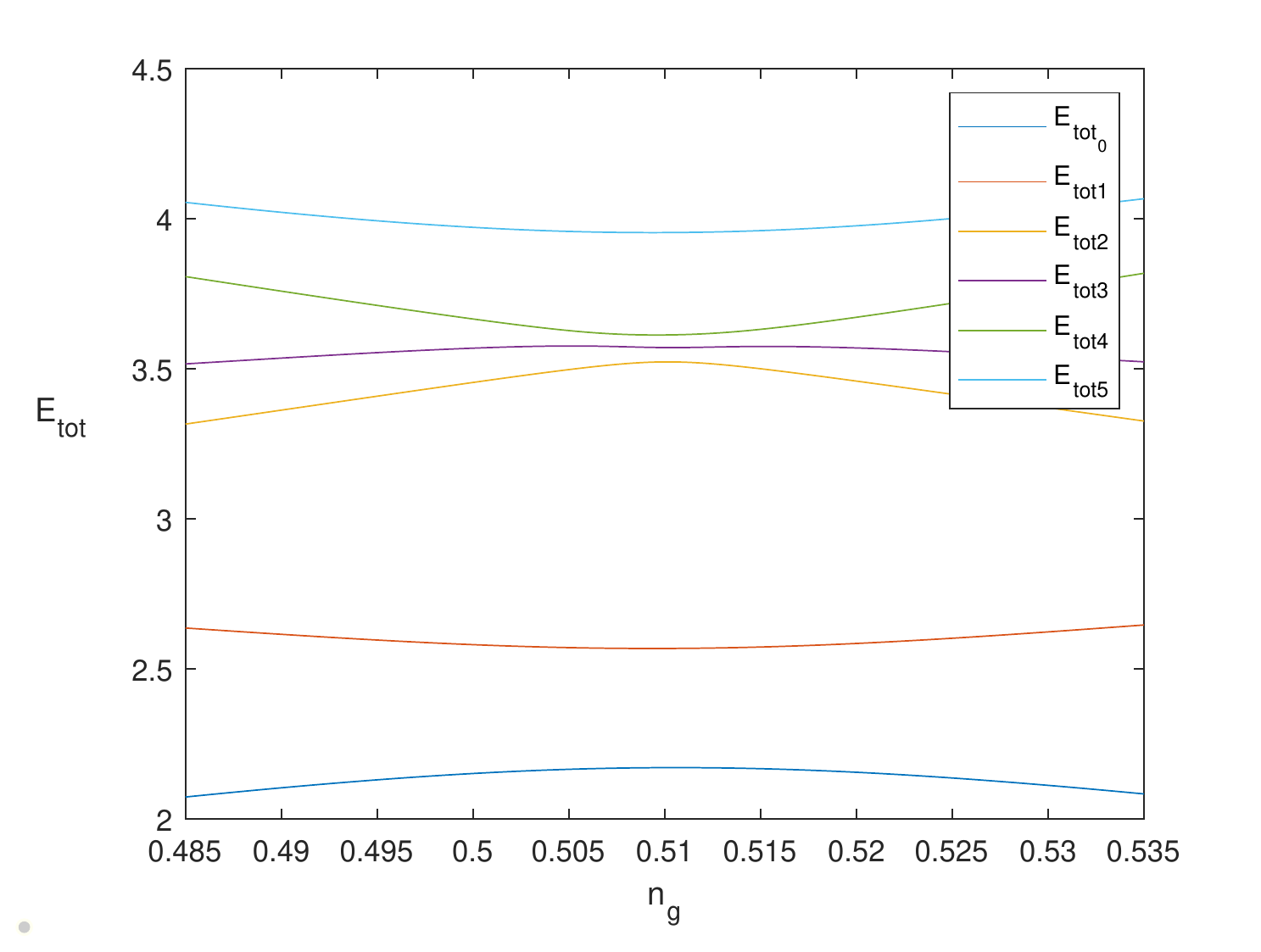}
			\caption{\centering $ E_J/ E_C=1 $}
		\end{subfigure}
		\begin{subfigure}{0.45\textwidth}
			\includegraphics[width=8.6 cm]{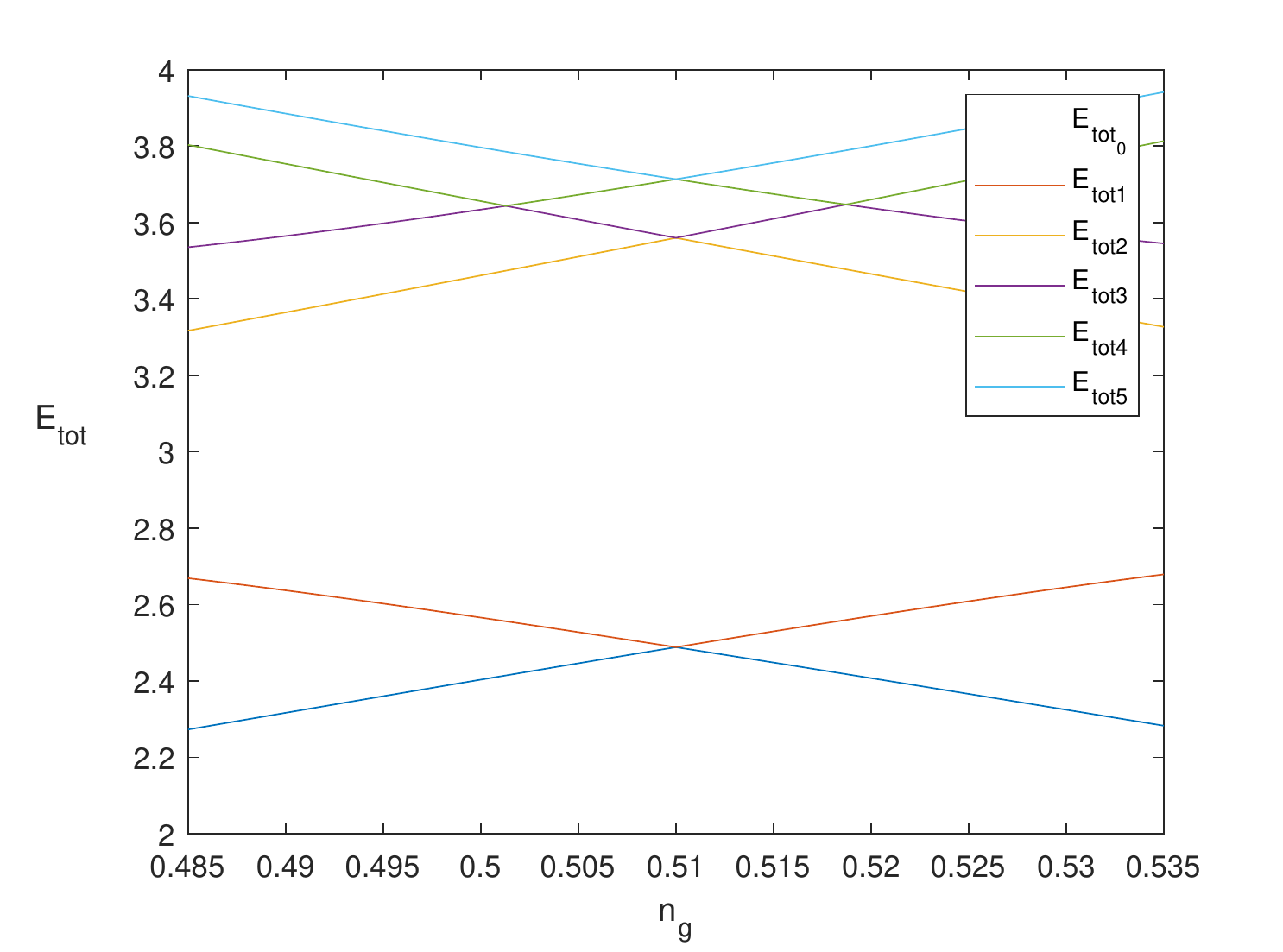}
			\caption{\centering $ E_J/ E_C=0.001 $}
		\end{subfigure}
	\end{center}
	\caption{Energy spectrum of $ E_{\text{tot}} $ vs $ n_g $ in three different cases. The relevant parameters we have used are $ E_C=1 $, $ h=E_C/5 $, $ \epsilon_C=10 $, $ N_g=0 $ and $ t_1=t_2=0.5 $. (a) When $ E_J/ E_C=1000 $, the energy gaps are invariant when the overall induced charge of the quantum dot is changed. (b)When $ E_J/ E_C=1 $, the minimum energy gaps occur at $ n_g=0.51 $. (c) When $ E_J/ E_C=0.001 $, the degenerate occurs at $ n_g=0.501,0.510,0.519 $.}
	\label{fig:ngvse}
\end{figure}

In this appendix, we diagonalize $  H_{\text{tot}} $ numerically to discuss the effect of the two additional gate-tunable valves on the two-MZM island in Sec.~\ref{sec:majorana}. Following the method in Ref.~\cite{PhysRevB.95.235305}, define the operator $ \Gamma_1=e^{i\phi/2}\gamma_1 $ and $ \Gamma_1=e^{i\phi/2}\gamma_{2N} $. For the decoupled MZM island $ H_C $, we consider the 7 lowest-energy state $ \ket{N_s,p_{12}} $ here:
\begin{eqnarray}\label{key}
\ket{0}&=&\ket{{N_S}=0;p_{12}}, \\\quad \ket{1}&=&\ket{{N_S}=1;-p_{12}}=\Gamma_1^\dagger\ket{0}\\
\ket{2}&=&\ket{{N_S}=-1;-p_{12}}=\Gamma_1\ket{0}\\
\ket{3}&=&\ket{{N_S}=2;-p_{12}}=\Gamma_1^\dagger\ket{1}\\
\ket{4}&=&\ket{{N_S}=-2;-p_{12}}=\Gamma_1\ket{2}\\
\ket{5}&=&\ket{{N_S}=3;-p_{12}}=\Gamma_1^\dagger\ket{3}\\
\ket{6}&=&\ket{{N_S}=-3;-p_{12}}=\Gamma_1\ket{4}
\end{eqnarray}
And the two states $ \ket{\tilde{n_f}} $ for quantum dot Hamiltonian $ H_{QD} $ are
\begin{eqnarray}\label{key}
\ket{\tilde{0}}&=&\ket{\tilde{n_f}=0};\\
\ket{\tilde{1}}&=&f^\dagger\ket{\tilde{0}}.
\end{eqnarray}
With the notation of $ \Gamma_i $, we can write $ H_\text{t,QD} $ in the form
\begin{equation}\label{key}
H_\text{t,QD}=-\frac{i}{2} f^\dagger(t_1\Gamma_{1}+t_2 \Gamma_{2N})+h.c.
\end{equation}
Noting that
\begin{equation}\label{key}
i\Gamma_1^\dagger\Gamma_{2N}=i\gamma_{1}\gamma_{2N}=p_{12}
\end{equation}
we get
\begin{eqnarray}\label{key}
\Gamma_{2N}\ket{0}=\Gamma_{2N}\Gamma_{1}\Gamma_{1}^\dagger\ket{0}=-ip_{12}\ket{2}\\
\Gamma_{2N}\ket{1}=\Gamma_{2N}\Gamma_{1}^\dagger\Gamma_{1}\ket{1}=-ip_{12}\ket{0}\\
\Gamma_{2N}\ket{2}=\Gamma_{2N}\Gamma_{1}\Gamma_{1}^\dagger\ket{2}=ip_{12}\ket{4}\\
\Gamma_{2N}\ket{3}=-\Gamma_{1}^\dagger\Gamma_{2N}\ket{1}=-ip_{12}\ket{1}\\
\Gamma_{2N}\ket{4}=\Gamma_{2N}\Gamma_{1}\Gamma_{1}\Gamma_{1}\Gamma_{1}^\dagger\ket{0}=-ip_{12}\ket{6}\\
\Gamma_{2N}\ket{5}=\Gamma_{2N}\Gamma_{1}^\dagger\Gamma_{1}^\dagger\Gamma_{1}^\dagger\ket{0}=-ip_{12}\ket{3}
\end{eqnarray}
\begin{eqnarray}
&&H_{\text{t,QD}}\sum_\mu\ket{\mu}\bra{\mu}\otimes \sum_\beta\ket{\tilde{\beta}}\bra{\tilde{\beta}}\nonumber\\
&&= (-\frac{i}{2} f^\dagger(t_1\Gamma_{1}+t_2 \Gamma_{2N}))\sum_\mu\ket{\mu}\bra{\mu}\otimes \sum_\beta\ket{\tilde{\beta}}\bra{\tilde{\beta}} +\text{h.c.}\nonumber\\
&&=-\frac{i}{2} (t_1\Gamma_{1}+t_2 \Gamma_{2N})(\ket{0}\bra{0}+\ket{1}\bra{1}+...+\ket{6}\bra{6})\otimes f^\dagger(\ket{\tilde{0}}\bra{\tilde{0}}+\ket{\tilde{1}}\bra{\tilde{1}})+\text{h.c.}\nonumber\\
&&=-\frac{i}{2}[ t_1(\ket{2}\bra{0}+\ket{0}\bra{1}+\ket{4}\bra{2}+\ket{1}\bra{3}+\ket{6}\bra{4}+\ket{3}\bra{5})\nonumber\\
&&~~~+t_2(-i p_{12}\ket{2}\bra{0}-ip_{12}\ket{0}\bra{1}+ip_{12}\ket{4}\bra{2}\nonumber\\
&&~~~-ip_{12}\ket{1}\bra{3}-ip_{12}\ket{6}\bra{4})+ip_{12}\ket{3}\bra{5}]
\otimes\ket{\tilde{1}}\bra{\tilde{0}}+\text{h.c.}
\end{eqnarray}
The Josephson energy term can be written as
\begin{equation}\label{key}
H_J=-\frac{E_J}{2}(\sum_{n\in \mathbb{Z}}\ket{N_S}\bra{N_S+1}+\ket{N_S+1}\bra{N_S})
\end{equation}
We can write the whole system Hamiltonian
\begin{equation}\label{key}
H_{\text{tot}}=H_C+H_J+H_{\text{QD}}^{\text{eff}}+H_{\text{t,QD}}.
\end{equation}
The energy eigenvalue can be obtained by diagonalizing the total Hamiltonian, as shown in Fig.~\ref{fig:ngvse}.
\bibliographystyle{utphys}
\bibliography{ref123}
\end{document}